# Performance Analysis of Novel Propellant Oxidizers using Molecular Modelling and Nozzle Flow Simulations

*Pujan Biswas[1], Sudarshan Kumar[1], Neeraj Kumbhakarna[2*]*

[1]*Department of Aerospace Engineering, Indian Institute of Technology Bombay, JVLR, Powai, Mumbai, 400076, India*
[2]*Department of Mechanical Engineering, Indian Institute of Technology Bombay, JVLR, Powai, Mumbai, 400076, India*
[*]*Corresponding Author Email: neeraj_k@iitb.ac.in*

**Abstract:** Search for alternate fuels for improvement in rocket engine performance is a topic of ever-growing interest and discussion in the research community. The primary target of this paper is to present novel compounds in view of their possible use as oxidizers in propulsion applications using molecular modeling calculations and supersonic flow simulations. Carbon-based heterocyclic compounds tend to have strained molecular structures leading to high heats of formation and energetic behavior. In the present work, molecular modeling calculations for molecules of 37 such potential propellant oxidizers are presented. Density functional theory (B3LYP) was employed for the geometry optimization of the proposed molecular structures using the 6-311++G(d,p) basis set. Heats of formation of the compounds were calculated using the molecular modeling results. Appropriate propellant compositions were considered with the proposed compounds as oxidizer components and Ideal specific impulse ($I_{vac,ideal}$*) was calculated for each composition assuming isentropic flow, computed using the NASA CEA software package. To predict the actual delivered specific impulse ($I_{vac,act}$*), supersonic nozzle flow simulations of equilibrium product gases of each propellant composition have been carried out using OpenFOAM. The standard k-epsilon turbulence model for compressible flows including rapid distortion theory (RDT) based compression term, has been employed. As the problem is inherently transient in nature, local time stepping (LTS) methodology has been further implemented to reach a steady-state solution. These simulations accounted for divergence losses, turbulence losses and boundary layer losses and gave a more realistic estimation of the specific impulse. It was observed that the $I_{vac,act}$* for all propellant compositions lie between 88% to 91% of the corresponding ideal value. The newly proposed oxidizers showed considerable improvement in propulsion performance as compared to ammonium perchlorate which is currently the most widely used oxidizer in solid rocket motors. The maximum improvement observed in Isp was 24 s.

## 1. Introduction

Ammonium perchlorate (AP) has been used extensively as an oxidizer in various solid propulsion systems. However, as research and environmental awareness progresses, several substitutes for AP are actively being considered. Trache et al. [1] mention certain oxidizers supporting this premise, which is further emphasized by Kettner et al. [2]. Kettner proposed discussions pertaining to molecules containing trinitromethyl groups as possible oxidizers due to their ability to impart higher oxygen balance. Additionally, Yu et al. [3] proposed novel energetic fluoride-based oxidizers with higher specific impulses. In our previous works on fuels [4] and oxidizers [5], we have attempted to list down some prospective novel compounds targeting propulsion applications. Detailed molecular optimizations to estimate the heats of formations were employed in Gaussian 09 [6] and $CO_2$ based oxygen balance numbers were calculated for the compounds. Preliminary specific impulse ($I_{sp}$) calculations were done using NASA CEA [7].

Although rocket nozzles experience high speed, high temperature flows, results in NASA CEA are based on isentropic and no flow loss assumptions. Thus, it is important for the scientific community to estimate the propulsive parameters in conditions that are close to reality. Thus, the objective of the current work is to provide more such compounds and estimate their propulsive performance under non-ideal rocket flow conditions.

Flow of combustion gases in a rocket nozzle is supersonic and several losses occur in this flow. Landsbaum et. al. [8] attempted to estimate the $I_{sp,vac}$* of propellants considering the losses due to two-phase flow, divergence, boundary layer, kinetics, nozzle submergence, and combustion inefficiency. They employed a proprietary solid performance prediction program (SPP) [9] to arrive at their results. One of the aims of the current paper is to provide an open-source, free-for-all methodology to estimate the delivered propulsive performance parameters for a given propellant composition. Bhat et. al. [10] presented a semiempirical approach to calculate $I_{sp,vac}$* of composite modified double base (CMDB) propellant formulations by utilizing calorimetric value (Cal-Val) of the propellant composition. But this paper does not mention the incorporated losses and nozzle configuration. Beckstead [11] developed a model to describe the combustion characteristics of composite solid propellants in detail but his work does not delve into the propulsive applications of the study, which is the primary focus of our work.

The propellants considered in this work are still in the development stage. With the advances in computational and numerical techniques, research community is moving towards estimating the non-ideal rocket performance based on simulations. Frem et. al. [12] attempted to predict the specific impulse of more than 165 compositions belonging to virtually all classes of propellants such as monopropellants, single-base, double-base, triple-base and composite modified double-base (CMDB) propellants, pseudo-propellants, composite propellants, liquid mono- and bipropellants, and finally hybrid propellants. They came up with their methodology based on the heat of reaction and the number of moles of gaseous reaction products per gram of propellant calculated according to ($H_2O$-$CO_2$) arbitrary decomposition assumption and constants derived from the ISPBKW code [13]. Although these predictions signify the importance of a computational estimation, their research considers only propellant combinations that do not contain any metal additives such as aluminum. The target of this current work is to keep such constraints at bay and generate a platform for all different varieties of compounds. Apte et. al. [14] conducted a time-resolved numerical analysis of combustion dynamics of double-base homogenous solid propellant in a rocket motor by using Large-Eddy Simulation (LES). They presented results of flame structure and turbulence-induced large-scale vortical motions, driving flow oscillations with distinct frequencies. Although Apte et. al. provided a piece of strong evidence for using computational fluid dynamics (CFD), they did not focus on the $I_{sp}$ characteristics of propellants. Further, Invigorito et. al. [15] presented results obtained through their simulation of the Penn State pre-burner combustor using CIRA of OpenFOAM [16] as their choice for a CFD tool. This paper brings out two important aspects; reduced computational cost and close relevance to experimental observations of the OpenFOAM opensource code.

Furthermore, this paper provides the much-needed insight to use Reynolds Averaged Navier Stokes (RANS) formulation in OpenFOAM to balance computational cost along with flexibility and fidelity of the model. Results in [15] implicate the reactingFoam solver implementing RANS – Menter Shear Stress Transport (SST) k-omega model for 9 species, supporting the use of OpenFOAM for reacting high temperature mixtures. While Invigorito et. al. discuss the combustion characteristics; our current work makes an attempt to predict the actual $I_{sp,vac}$* for our proposed compounds using a supersonic nozzle-flow simulation utility in OpenFOAM. The actual $I_{sp,vac}$*, unlike the ideal one predicted by NASA CEA calculations, accounts for dissipative and turbulence losses primarily in the nozzle. Thus, losses arising due to boundary layer, turbulence and divergence are included and a more realistic estimation of propulsion is obtained. With regard to combustion, equilibrium product composition obtained from NASA CEA has been considered in the combustion chamber without any further chemical change while flowing through the nozzle. This was done because combustion losses are a part of previous studies carried out by various researchers (e.g. Ecker et. al. [17] and Apte et. al. [14]). Table 1 lists the envisaged molecules which are the object of current study.

## 2. Methods

### *2.1 Molecular modelling*

Molecular structures were optimized using quantum mechanics-based calculations by employing Gaussian 09. We have used B3LYP/6-311++G(d,p) level of theory consisting of triple split valence basis set with additional polarized functions [18]. Our choice of the aforementioned level of theory is to maintain a balance between computational time and accuracy. The detailed procedure of all calculations is described in our previous work [4]. The obtained optimized structures, oxygen balance numbers and heats of formation of the proposed 37 compounds are listed in Table 1.

Table 1: Molecular properties of proposed oxidizers

| Sr. No. | Name | Molecular Formula | Structure | HoF (KJ/mol) | $OB_{CO2}$ (%) |
|---|---|---|---|---|---|
| 1 | S2-F | $C_2H_4N_8O_7$ | 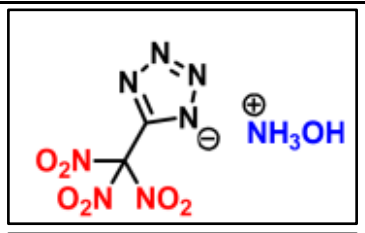 | 741.9487 | 6.35 |
| 2 | S2-E | $C_2H_4N_8O_6$ | 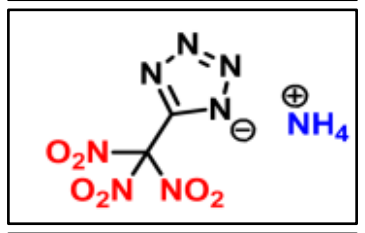 | 700.0669 | 0 |
| 3 | S2-2* | $C_2N_8O_9$ | 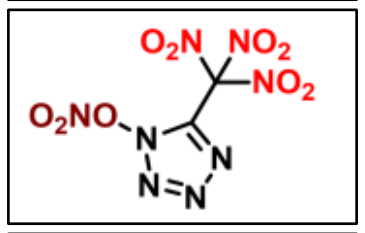 | 478.2312 | 28.6 |
| 4 | S10-5 | $CN_6O_4$ | 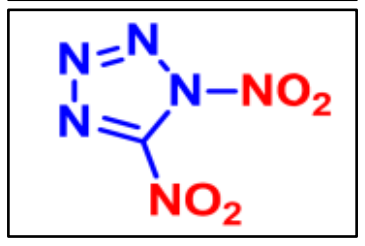 | 477.8128 | 20.0 |
| 5 | S1-3 | $C_2N_{12}O_6$ | 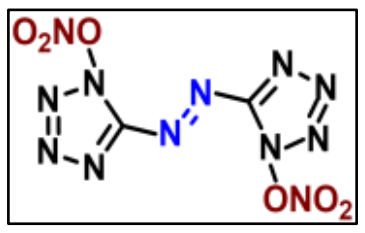 | 1076.711 | 11.2 |
| 6 | S2-4 | $C_2HN_9O_8$ | 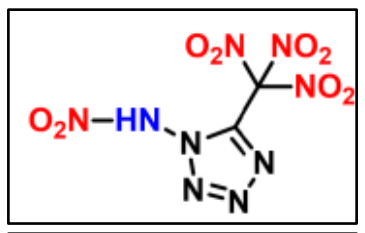 | 555.3005 | 20.1 |
| 7 | S2-5 | $C_4N_{16}O_{12}$ | 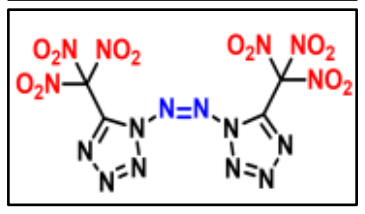 | 1271.978 | 13.8 |
| 8 | S1-4 | $C_2N_8O_8$ | 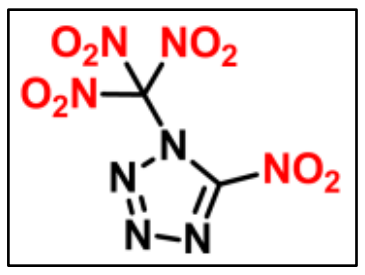 | 481.2855 | 24.2 |
| 9 | S1-5 | $C_2N_9O_8$ | 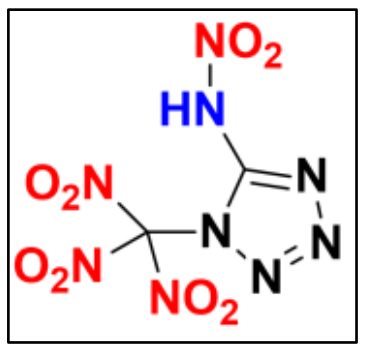 | 490.1138 | 20.1 |
| 10 | S2-1 | $C_2HN_7O_7$ | 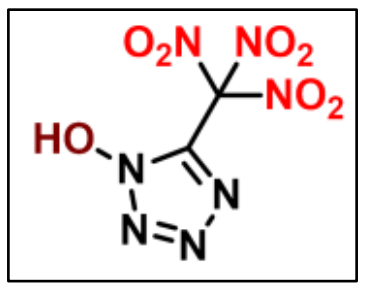 | 387.9405 | 17.0 |
| 11 | S6-3* | $CHN_5O_8$ | 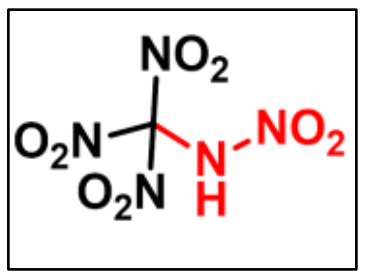 | -63.5968 | 41.71 |
| 12 | S1-1 | $CHN_7O_4$ | 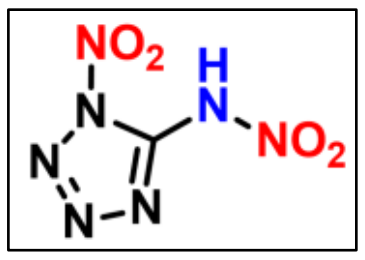 | 480.8671 | 13.7 |
| 13 | S6-2 | $C_2HN_7O_6$ | 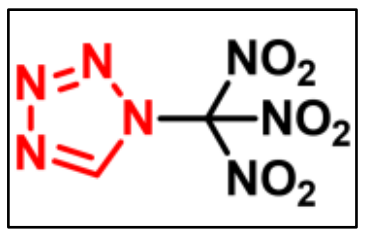 | 423.4208 | 10.96 |
| 14 | S10-3 | $C_2N_{12}O_6$ | 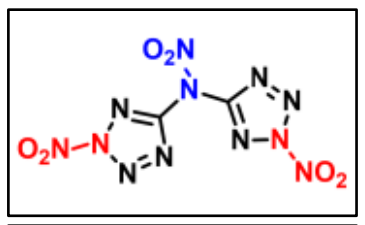 | 961.7761 | 11.11 |
| 15 | S6-1 | $C_6H_6N_{12}O_{18}$ | 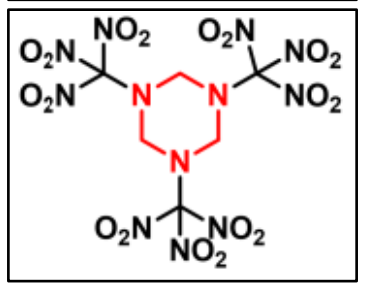 | 288.8215 | 8.99 |
| 16 | S12-1 | $C_4N_{10}O_{15}$ | 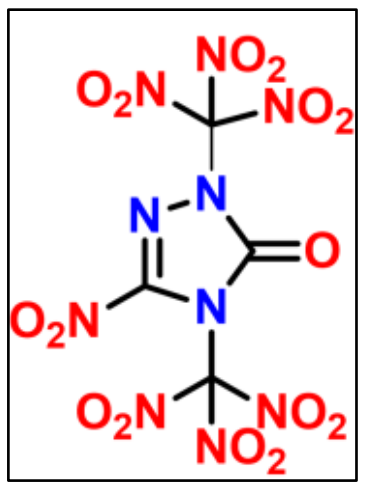 | 238.6554 | 26.20 |
| 17 | S10-2 | $C_2H_4N_4O_8$ | 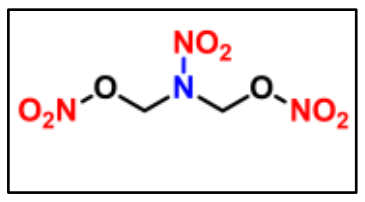 | -171.67 | 15.09 |
| 18 | S12-2 | $C_4HN_9O_{13}$ | 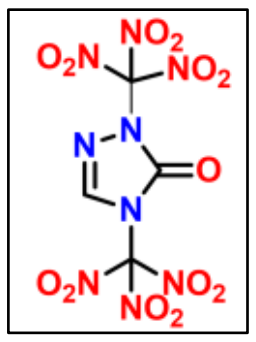 | 161.2095 | 18.80 |
| 19 | S11-2 | $C_6N_{16}O_{14}$ | 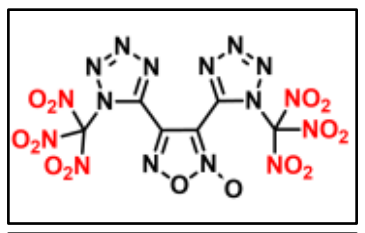 | 1156.29 | 6.15 |
| 20 | S9-2 | $C_6N_{12}O_{16}$ | 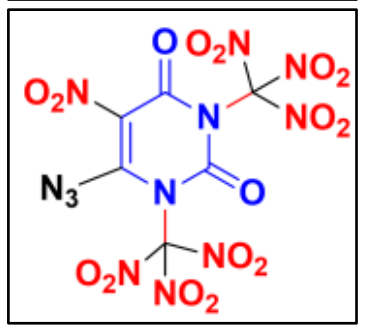 | -67.3206 | 12.90 |
| 21 | S4-3’ | $C_4N_6O_{12}$ | 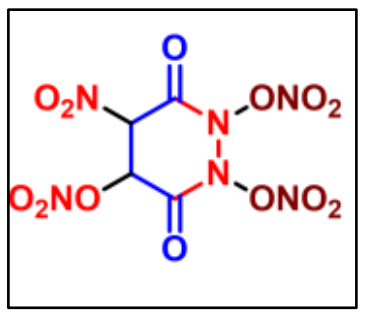 | -200.121 | 18.72 |
| 22 | S3-4 | $C_4N_{10}O_{14}$ | 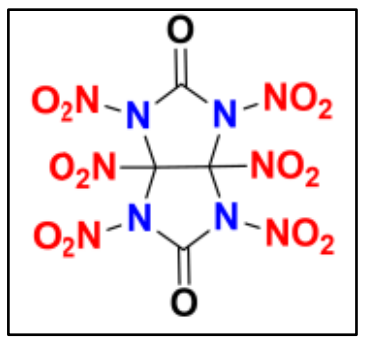 | 162.5902 | 23.29 |
| 23 | S8-1 | $C_6HN_{13}O_{14}$ | 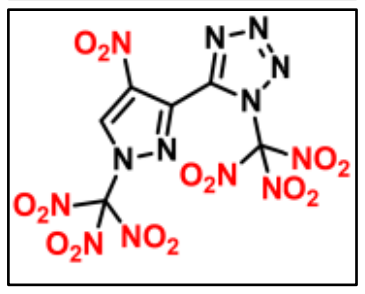 | 753.7476 | 5.01 |
| 24 | S11-1 | $C_6H_2N_{18}O_{14}$ | 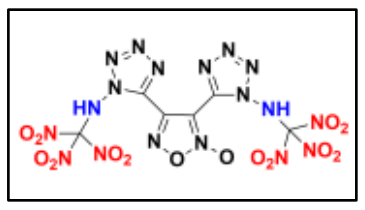 | 1277.71 | 2.90 |
| 25 | S3-1 | $C_6H_2N_{12}O_{18}$ | 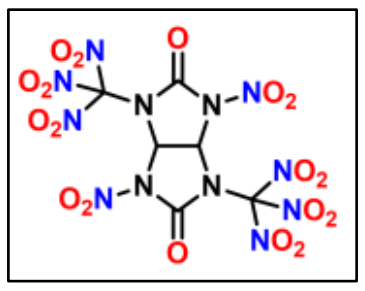 | 77.6132 | 15.09 |
| 26 | S10-1 | $CH_2N_4O_5$ | 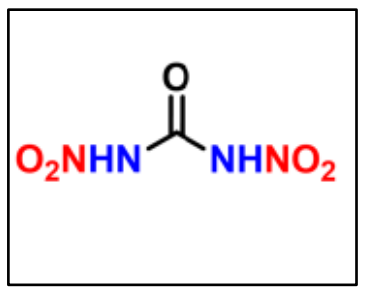 | -94.0563 | 21.33 |
| 27 | S1-2 | $C_2N_{12}O_4$ | 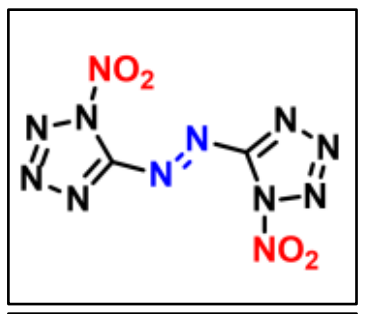 | 1163.194 | 0 |
| 28 | S4-2 | $C_4N_6O_{12}$ | 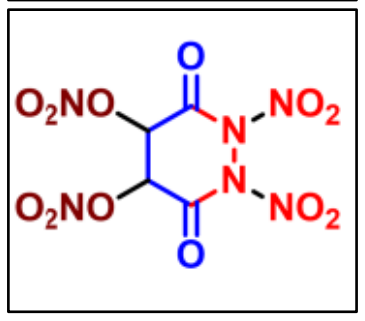 | -152.172 | 14.72 |
| 29 | S12-3 | $C_3HN_7O_9$ | 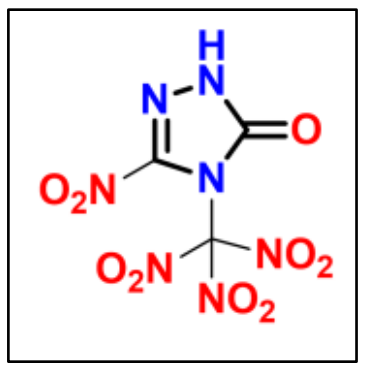 | 92.38272 | 14.3 |
| 30 | S9-4 | $C_5N_{12}O_{15}$ | 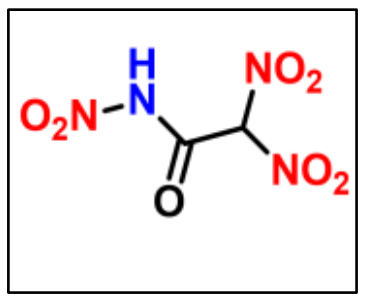 | -132.256 | 16.48 |
| 31 | S3-1’ | $C_5N_{12}O_{15}$ | 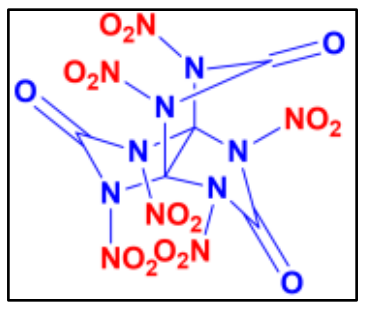 | 235.9776 | 17.09 |
| 32 | S4-1’ | $C_4N_6O_{11}$ | 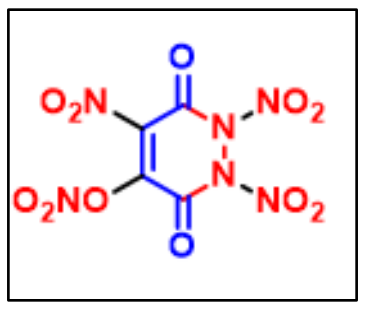 | -8.1588 | 15.6 |
| 33 | S9-1 | $C_6HN_9O_{16}$ | 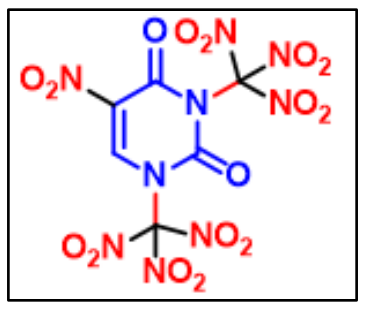 | -8.70272 | 12.3 |
| 34 | S5-2 | $C_7H_4N_{10}O_{20}$ | 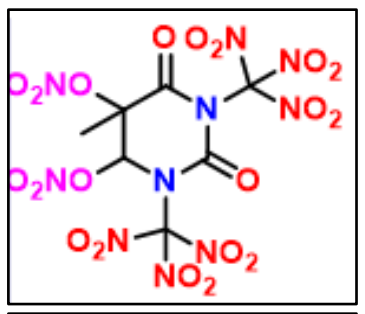 | -227.568 | 11.7 |
| 35 | S5-1’ | $C_7H_4N_{10}O_{18}$ | 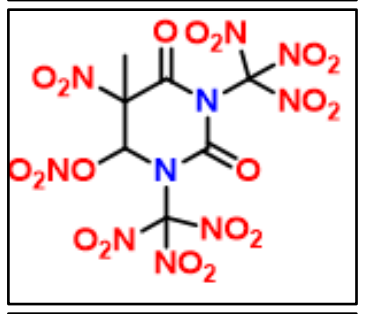 | -107.947 | 9.0 |
| 36 | S4-4* | $C_6H_4N_8O_{16}$ | 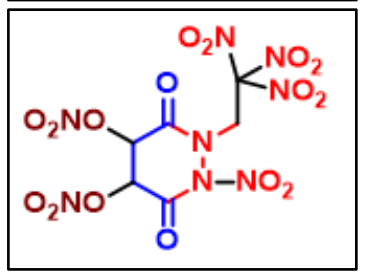 | -166.732 | 7.2 |
| 37 | S8-2 | $C_6H_2N_{12}O_{12}$ | 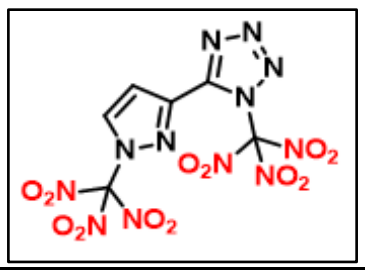 | 712.8281 | -3.69 |

*2.2 Nozzle Flow Simulations*

As indicated in [15], the use of reactingFoam solver of OpenFOAM is suitable for reacting flow simulations. In our case, the equilibrium combustion products obtained from NASA CEA calculations are considered to enter the nozzle from the combustion chamber. The governing equations of continuity, momentum, energy and transport for the system considered are:

$$\frac{\partial \rho}{\partial t} + \nabla \cdot (\rho \boldsymbol{U}) = 0 \tag{1}$$

$$\frac{\partial (\rho \boldsymbol{U})}{\partial t} + \nabla \cdot (\rho \boldsymbol{U}\boldsymbol{U}) = -\nabla p + \nabla \cdot \tau + \rho \boldsymbol{g}; \tau = \mu \left[ (\nabla \boldsymbol{U} + \nabla \boldsymbol{U}^T) - \frac{2}{3} \nabla \cdot \boldsymbol{U} I \right] \tag{2}$$

$$\frac{\partial (\rho h)}{\partial t} + \nabla(\rho \boldsymbol{U} h) + \frac{\partial (\rho K)}{\partial t} + \nabla(\rho U K) - \frac{\partial p}{\partial t} = \nabla \alpha_{eff} \nabla h \tag{3}$$

$$\frac{\partial (\rho Y_s)}{\partial t} + \nabla \cdot (\rho \boldsymbol{U} Y_s) = \nabla \mu_{eff} \nabla Y_s \tag{4}$$

Here, $\rho$ denotes the density, $\boldsymbol{U}$ is the velocity, $p$ is the pressure, $h = \sum_s Y_s \int C_{p,s} dT$ is the enthalpy, $T$ is the temperature, $Y_s$ is the species mass fraction, $C_{p,s}$ is the heat capacity for species $s$ in the mixture and $\alpha_{eff}$ is the effective thermal diffusivity, $\mu_{eff}$ is the effective dynamic viscosity based on turbulence model and $K$ is the kinematic energy. All thermodynamic properties are calculated using the coefficients from JANAF thermochemical tables. The temperature dependencies of the pure species specific heat capacities are fitted by Equation 5 and considering 7 terms with $q_i = -2$ to 4 as per Least-Squares Fit by NASA.

$$\frac{C_p^0(T)}{R} = \sum_{i=1}^{r} a_i T^{q_i} \tag{5}$$

The extended details are explained in the works of Dmitry et. al. [19], Mousavi [20] and Bonart et. al. [21]. Following up on their review and research, this work also employs the RANS k-epsilon model discretized by the Euler method. OpenFOAM employs the k-epsilon model for compressible flows including rapid distortion theory-based compression term. The Sutherland transport model, JANAF thermodynamic model and sensible enthalpies were used to calculate the temperature, pressure and velocity of the perfect gas composition at the various points of the geometry.

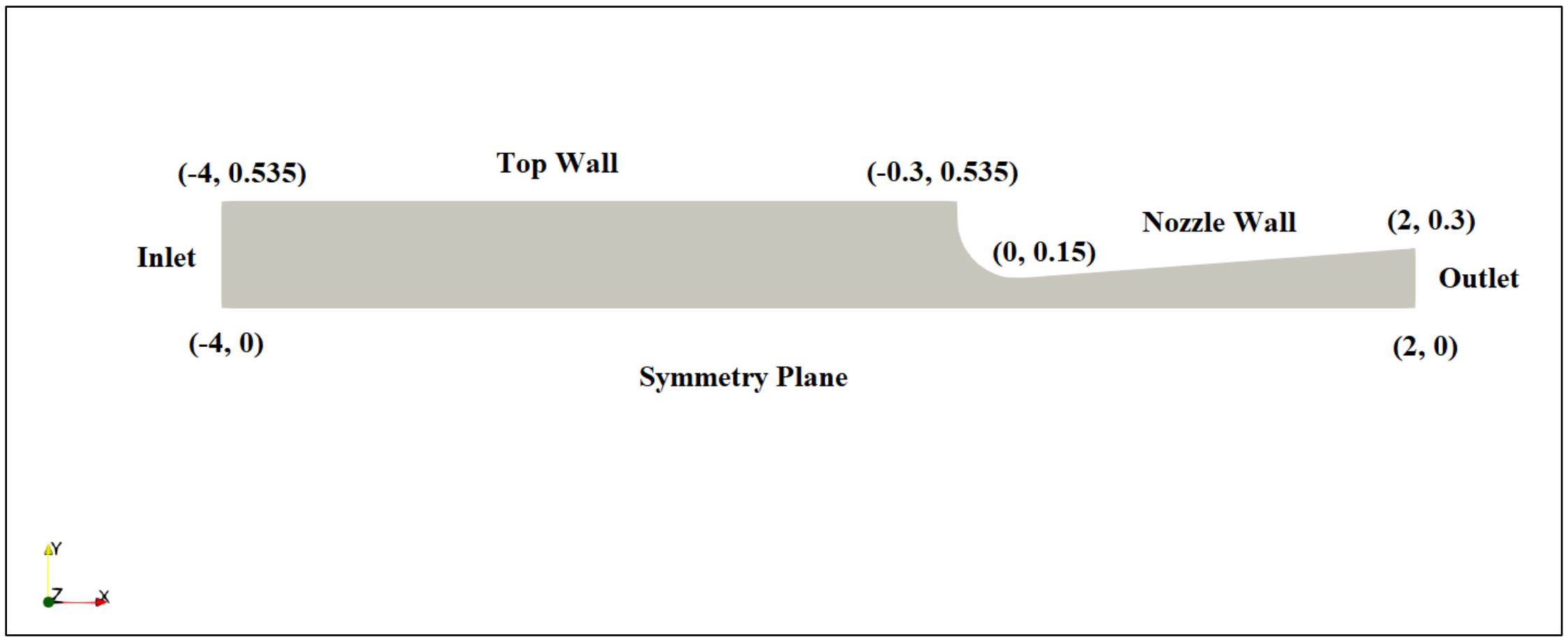

Figure 1. Nozzle diagram

Figure 1 shows the supersonic nozzle with $A_e/A_t = 4$ in consideration for the flow post combustion chamber. This specific nozzle shape was chosen to keep the computational effort lower while considering divergence and turbulence losses effectively. All the units are in meters. The total nozzle length is 2.6 m and 3.4 m of buffer length is provided post combustion. Olivera et. al. [22] showed in their work the correlation two-dimensional nozzle CFD with experimental results. Taking cue from them, we have also simulated flow through a two-dimensional nozzle to keep the analysis computationally feasible. The geometry has planar symmetry in the Z-X plane with a grid depth of 0.01 m in the Z-direction. The acceleration due to gravity (9.81 $m/s^2$) is applied in the positive X-direction as expected.

The combustion chamber stagnant pressure was set to 68.9 bar which is a typical rocket chamber pressure. Stagnation temperature was set to the equilibrium temperature of the propellant product mixture. The composition considered was 3% DOA, 11% HTPB, and 86% metal infused oxidizer (Al + Oxidizer). Within the 86% the relative amounts of Al and oxidizer were chosen using NASA CEA calculations such that the propellant composition gives the maximum $I_{sp,vac}$ and this maximum value is denoted by $I_{sp,vac}$*. No slip velocity boundary condition was set for the nozzle and top walls (as shown in Figure 1). Similarly, zero gradient pressure and temperature boundary conditions was set for the nozzle and top walls implying adiabatic wall. The velocity at the inlet of geometry was set to be automatically calculated based on the stagnant pressure and temperature. Similarly, no constraints were set for the outlet of the nozzle geometry and time-derived analysis would provide the corresponding values. Initially, the nozzle was nitrogen purged while the initial temperatures and pressures were estimated to the corresponding ideal values from NASA CEA. Rocket flows are inherently transient in nature but to calculate $I_{sp,vac}$*, steady state values need to be estimated. OpenFOAM reactingFoam solver takes into consideration only the transient states. Therefore, a method called as Local Time Stepping (LTS) has been implemented which predicts the steady state values when the time derivative term numerically tends to zero in the RANS equations. Once the pseudo steady state is obtained from the transient solver, the parameters are then fed in the LTS solver (discretized using local Euler method) which arrives at the steady state solution through iterations. The mathematical formulation of this procedure has been documented in detail by Kelley et. al [23].

One of the important aspects of CFD is to construct the grid and test for grid independent results.

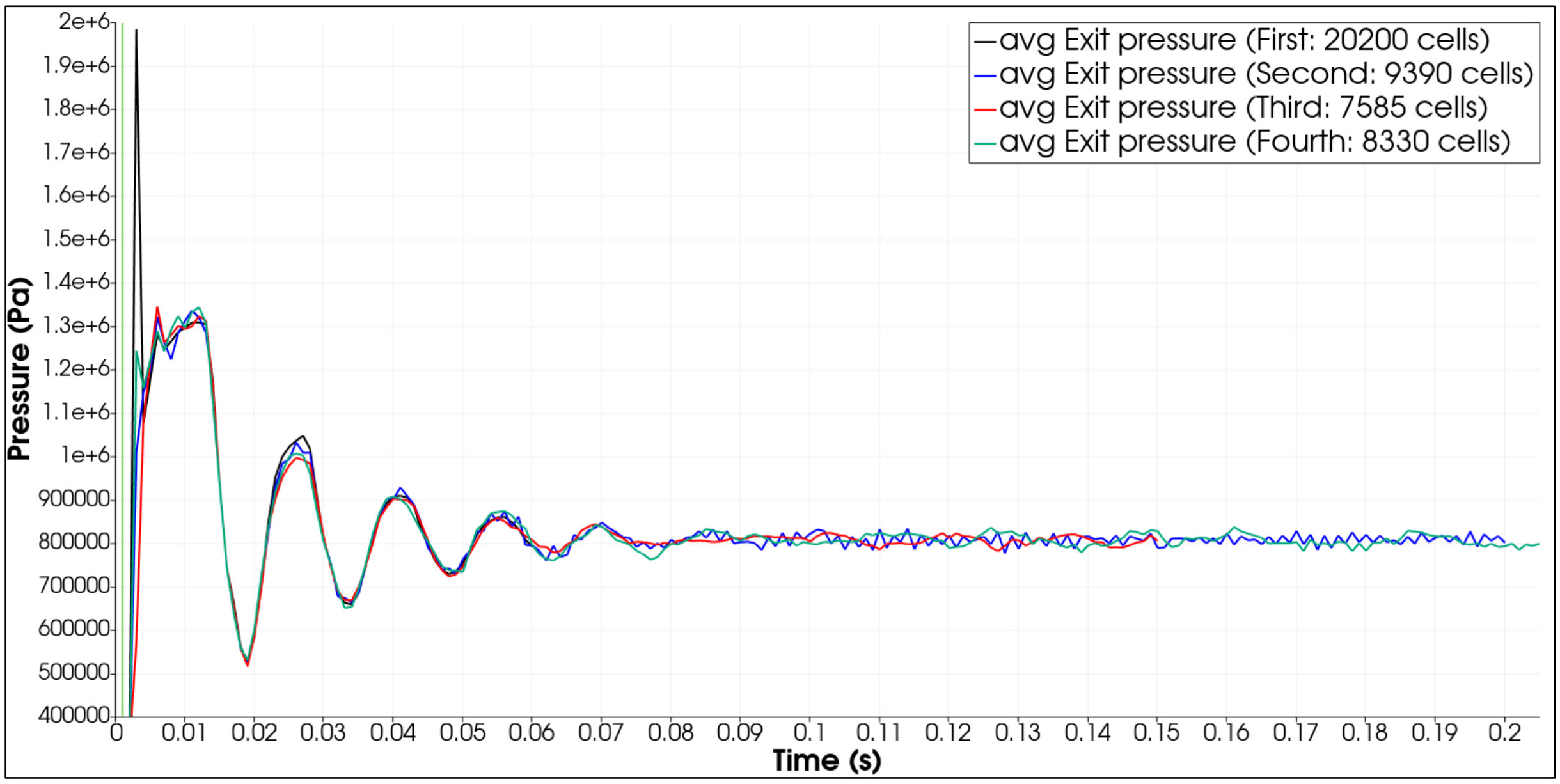

Figure 2. Average pressure covergence for 4 different grid sizes at the outlet

Figure 2. is the plot of the average pressure at the outlet for 4 different numbers of total grid cells. In Figure 2, First implies 20200 cells, Second implies 9390 cells, Third implies 7585 cells and Fourth implies 8330 cells. It is evident from Figure 2 that the transient state results converge to approximately the same value.

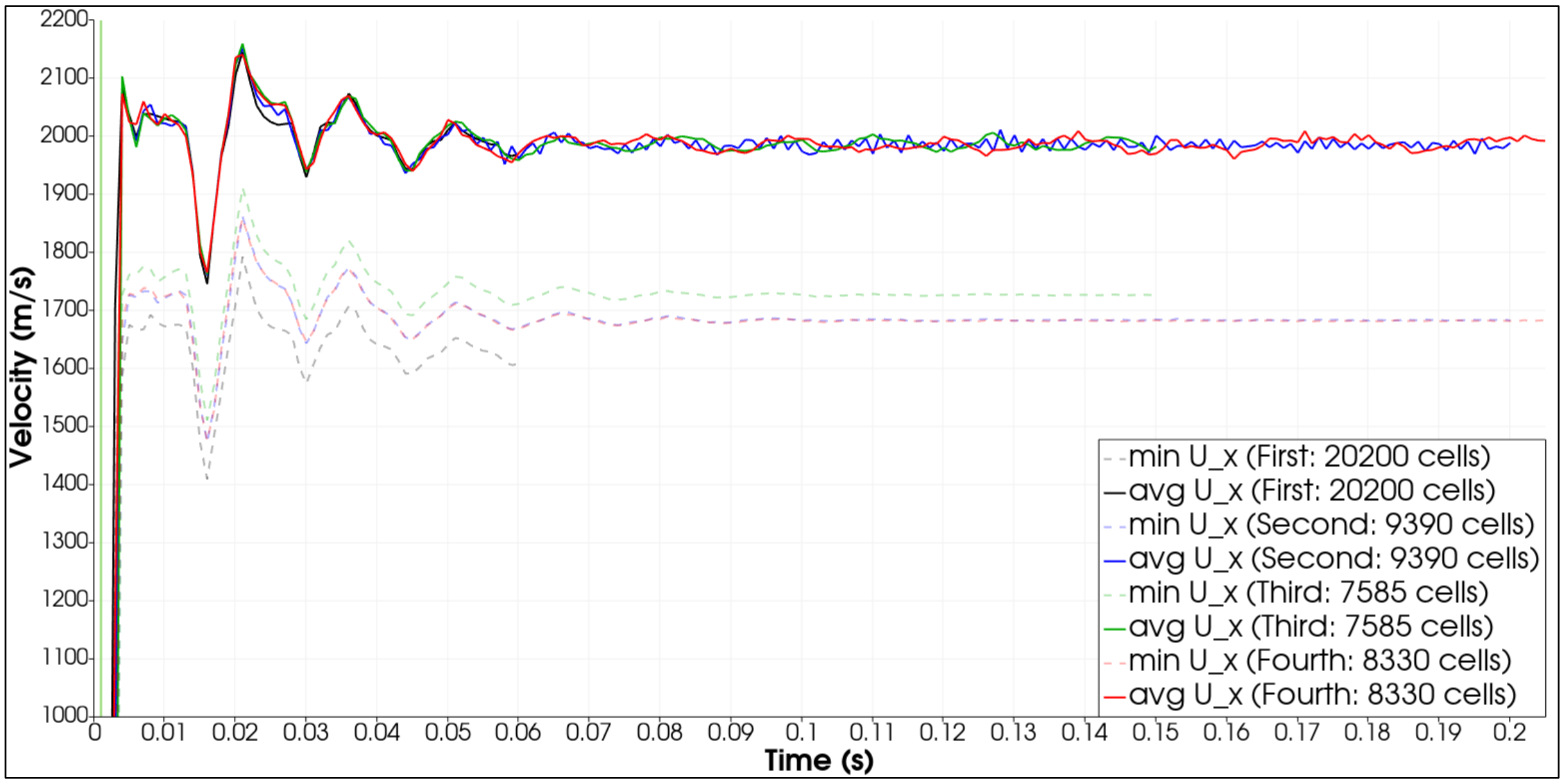

Figure 3. X-direction velocity at the nozzle exit (average and minimum)

Figure 3 gives interesting insights about the grid selection. The minimum velocities are different for the grids but the average velocities are same. This is because the minimum velocity at the outlet occurs at the nozzle wall grid center. A lower nozzle exit minimum velocity implies a denser grid as it is closer to the nozzle wall. However, higher computational time is required for finer grid sizes, in this case by a factor of ~4 between First and Third grid designs. Since the minimum grid size is also determined by the Courant number (set as 0.95 {<1}), there is a limitation to the fineness of the grid if a feasible timestep has to be considered. As, we have the same average velocities, irrespective of grid size, a coarser grid can be selected. The simulation in the fourth grid runs in approximately same computational time as compared to the third grid but comes closer in accuracy. This is due to the fact that it has more cells (30) in the Y-direction compared to the third (25). All results henceforth are shown for the fourth grid design. Figure 4 and Figure 5 showcase the fourth grid design wherein. Cells have been graded in 10:1 ratio up to the nozzle throat to obtain fineness in nozzle throat region. A fine grid size is similarly selected in the diverging nozzle section where divergence losses are expected to be maximum as compared to post combustion chamber zone.

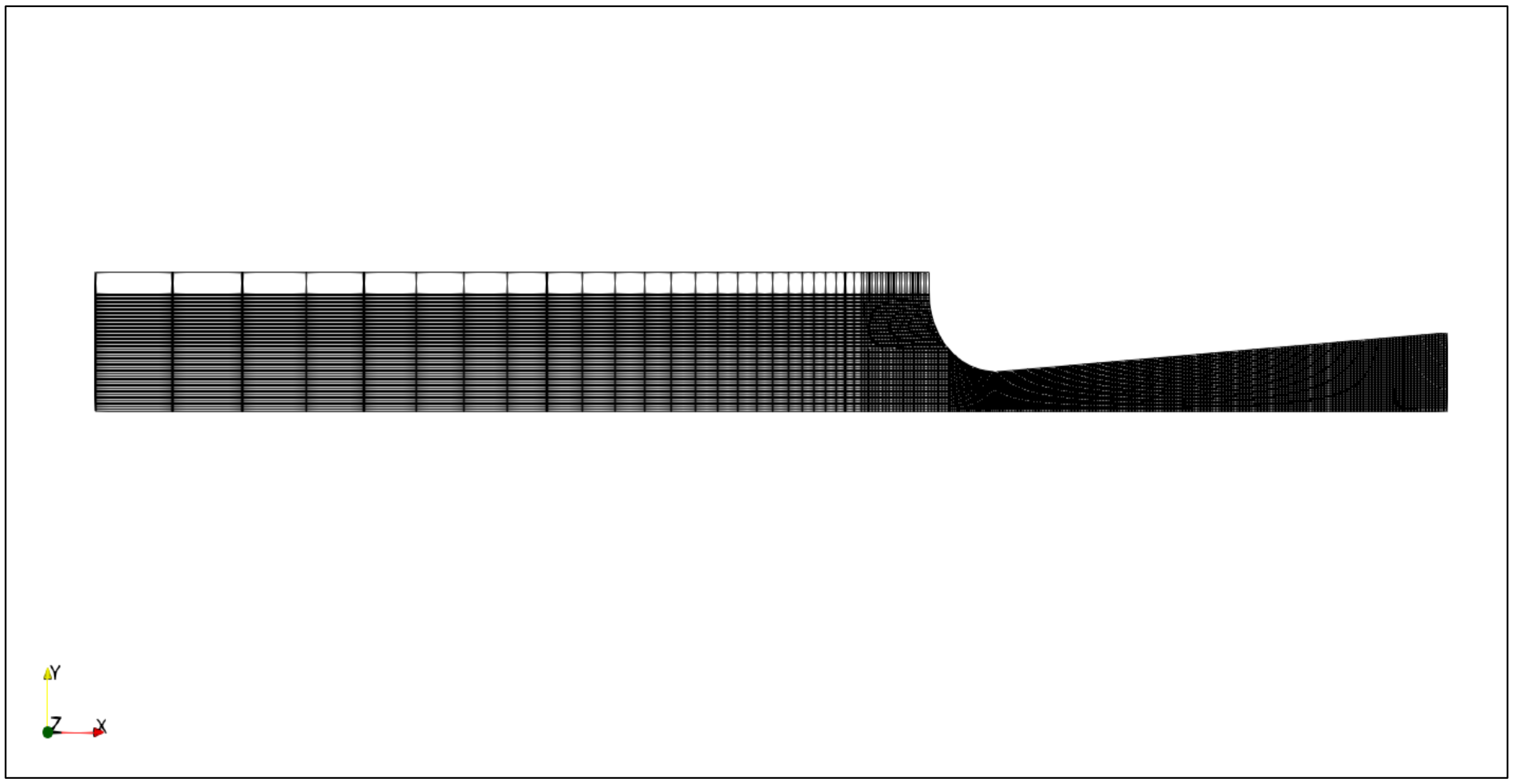

Figure 4. Nozzle geometry and mesh

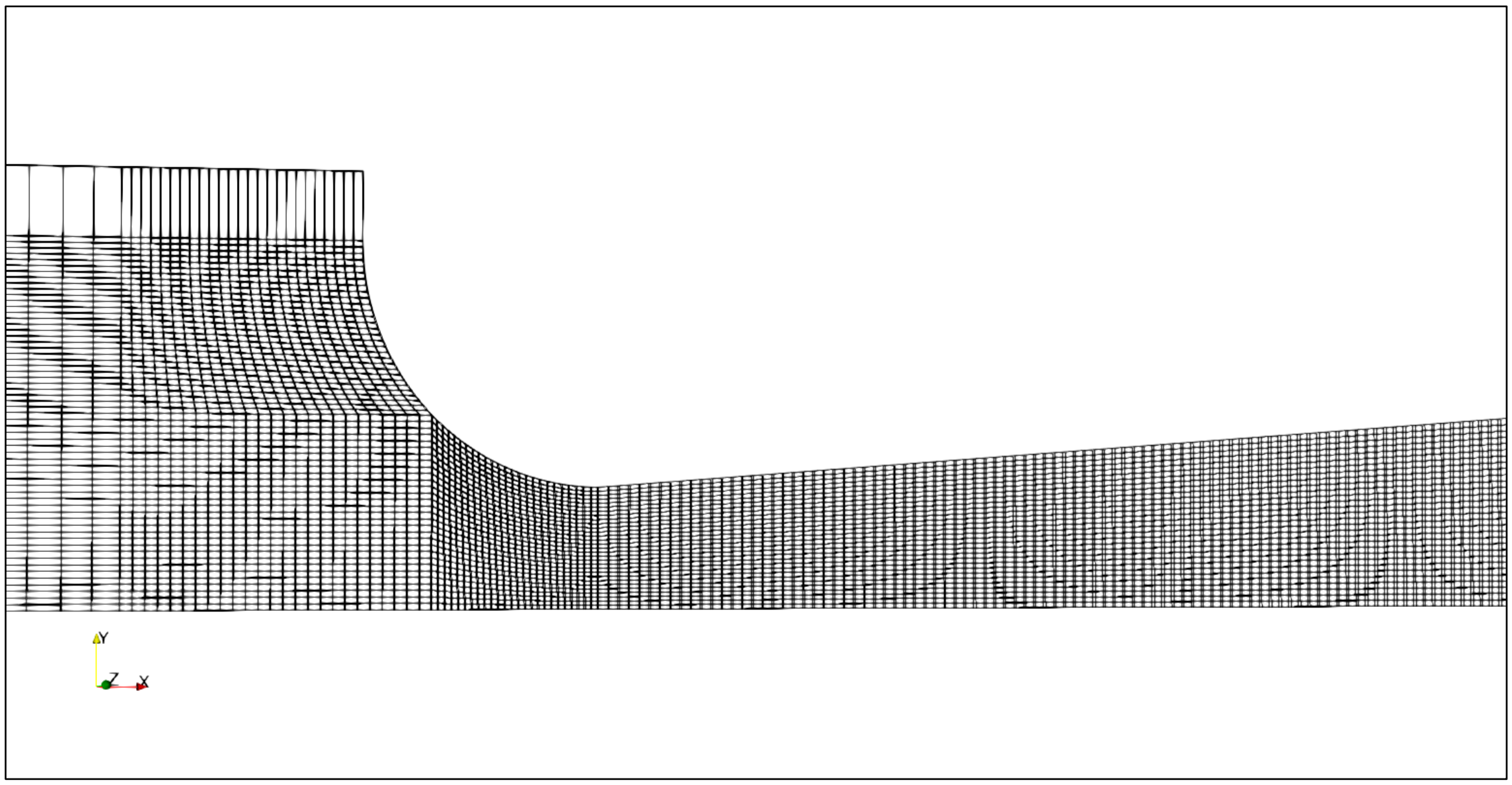

Figure 5. Panned-in view of the nozzle geometry and mesh

## 3. Results and Discussion

Figure 6 and Figure 7 are provided to show the performance in terms of exit average pressure- for the case in which oxidizer is AP. Figure 6 shows the transient solver solution. Similarly, Figure 7 shows the LTS solver solution which is evidence for a steady solution being reached.

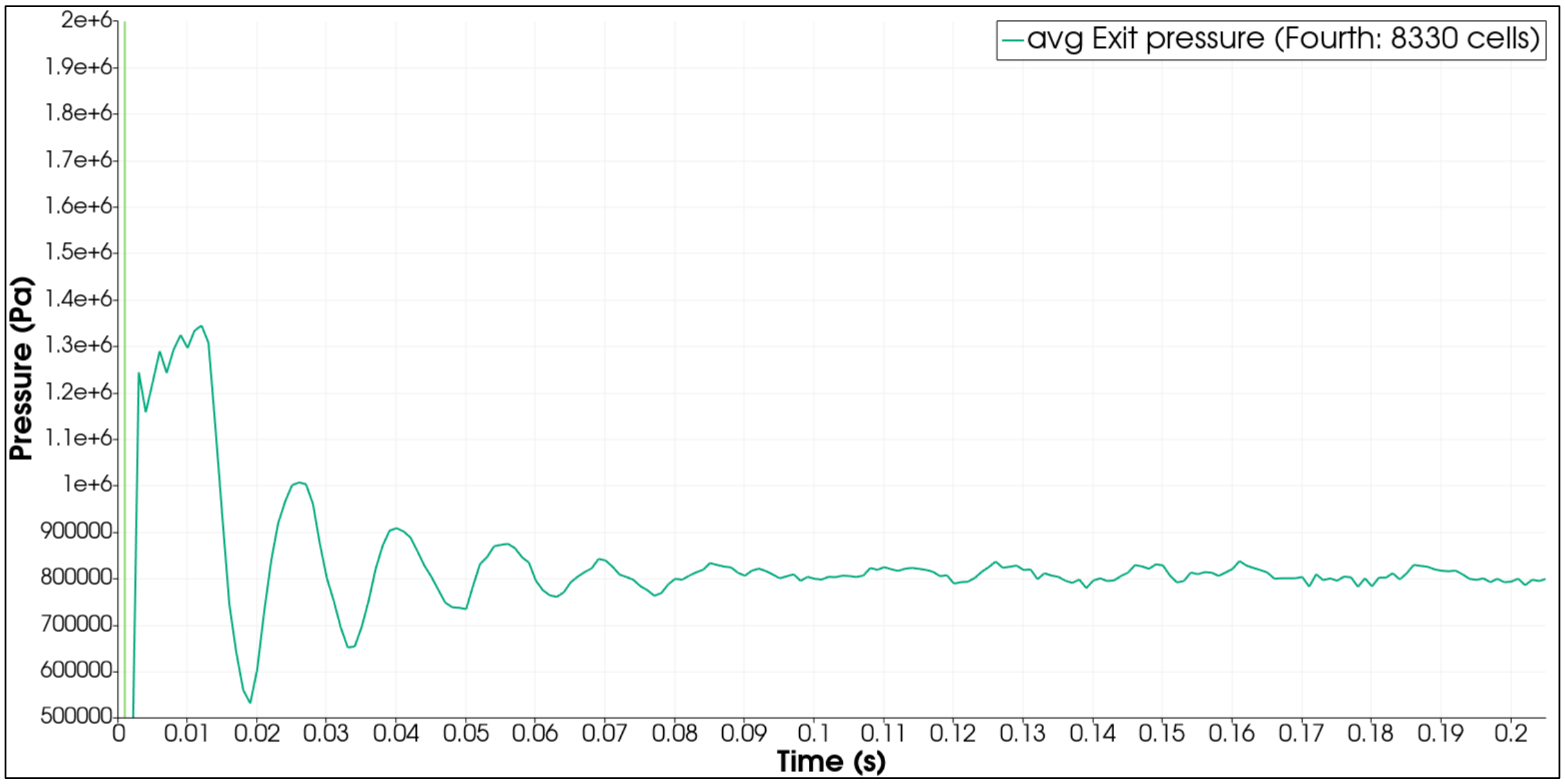

Figure 6. Average pressure at nozzle exit through transient solver phase for AP

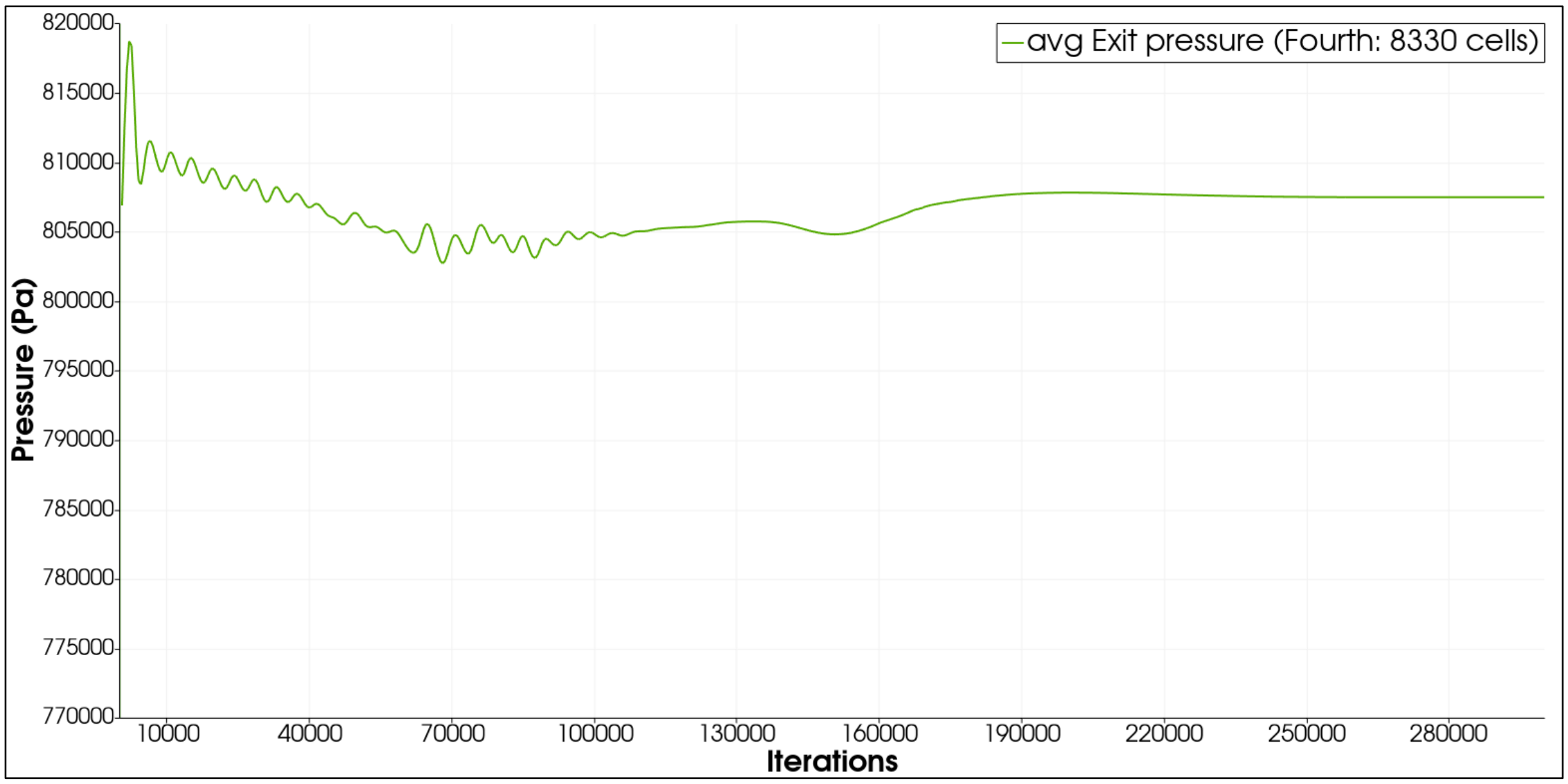

Figure 7. Average pressure at nozzle exit through LTS solver phase for AP

Similar results have been observed for all the other compounds. From these exit values, the starred specific impulse is calculated using Equations 6,7 and 8:

$$I_{sp}^{*} = \frac{1}{No.of cells\ at\ exit} \Sigma I_{sp,unit\ exit\ cell}^{*} \tag{6}$$

$$I_{sp,unit\ exit\ cell}^{*} = \frac{v_{unit\ exit\ cell}}{g} + \frac{(p_{unit\ exit\ cell}^{e} - p_0)}{g\left(\frac{\dot{m}_{unit\ exit\ cell}}{A_{unit\ exit\ cell}^{e}}\right)} \tag{7}$$

$$\frac{\dot{m}_{unit\ exit\ cell}}{A^{e}_{unit\ exit\ cell}} = \frac{p^{e}_{unit\ exit\ cell} * v_{unit\ exit\ cell}}{R * T_{unit\ exit\ cell}} \text{ where } R = \frac{R_{universal}}{MW_{exhaust\ gases}} \tag{8}$$

Thus, resulting non-ideal $I_{sp,vac}$* values for various propellant compositions along with the exact composition of the compounds the ideal $I_{sp,vac}$* is enlisted in Table 2. Efficiency in Table 2 represents the percentage ratio of the non-ideal $I_{sp,vac}$* to ideal $I_{sp,vac}$*.

Table 2. Comparative results between NASA CEA and OpenFOAM simulations

| **Sr. No.** | **Oxidizer (mass %)** | **Oxidizer (mass %)** | **Aluminum (mass %)** | **Ideal $I_{sp,vac}^*$ (s)** | **Non-ideal $I_{sp,vac}^*$(s)** | **Efficiency (%)** |
|---|---|---|---|---|---|---|
| 1 | S1-1 | 80 | 6 | 281.17 | 249.63 | 88.78 |
| 2 | S1-2 | 86 | 0 | 276.61 | 246.49 | 89.11 |
| 3 | S1-3 | 83 | 3 | 285.44 | 250.91 | 87.90 |
| 4 | S1-4 | 77 | 9 | 281.82 | 249.61 | 88.57 |
| 5 | S1-5 | 77 | 9 | 278.54 | 247.41 | 88.82 |
| 6 | S10-1 | 71 | 15 | 266.77 | 241.61 | 90.57 |
| 7 | S10-2 | 70 | 16 | 274.18 | 247.71 | 90.35 |
| 8 | S10-3 | 83 | 3 | 279.33 | 247.14 | 88.48 |
| 9 | S10-5 | 80 | 6 | 286.62 | 252.50 | 88.10 |
| 10 | S11-1 | 82 | 4 | 273.77 | 242.31 | 88.51 |
| 11 | S11-2 | 81 | 5 | 275.12 | 243.85 | 88.64 |
| 12 | S12-1 | 74 | 12 | 274.68 | 245.24 | 89.28 |
| 13 | S12-2 | 74 | 12 | 271.93 | 243.30 | 89.47 |
| 14 | S12-3 | 75 | 11 | 267.05 | 239.55 | 89.70 |
| 15 | S2-1 | 77 | 9 | 280.30 | 247.41 | 88.27 |
| 16 | S2-2* | 77 | 9 | 283.27 | 250.55 | 88.45 |
| 17 | S2-4 | 77 | 9 | 282.81 | 250.79 | 88.68 |
| 18 | S2-5 | 81 | 5 | 283.24 | 250.40 | 88.40 |
| 19 | S2-E | 80 | 6 | 292.34 | 258.83 | 88.53 |
| 20 | S2-F | 81 | 5 | 297.73 | 263.35 | 88.45 |
| 21 | S3-1 | 74 | 12 | 268.35 | 240.62 | 89.67 |
| 22 | S3-1’ | 75 | 11 | 267.07 | 238.97 | 89.48 |
| 23 | S3-4 | 74 | 12 | 270.91 | 238.98 | 88.21 |
| 24 | S4-1’ | 74 | 12 | 266.91 | 239.12 | 89.59 |
| 25 | S4-2 | 73 | 13 | 265.07 | 238.50 | 89.98 |
| 26 | S4-3’ | 72 | 14 | 263.29 | 236.73 | 89.91 |
| 27 | S4-4* | 74 | 12 | 267.24 | 240.31 | 89.92 |
| 28 | S5-1’ | 75 | 11 | 265.72 | 238.74 | 89.85 |
| 29 | S5-2 | 73 | 13 | 267.66 | 240.73 | 89.94 |
| 30 | S6-1 | 75 | 11 | 276.20 | 247.53 | 89.62 |
| 31 | S6-2 | 79 | 7 | 277.74 | 247.24 | 89.02 |
| 32 | S6-3* | 70 | 16 | 272.83 | 245.63 | 90.03 |
| 33 | S8-1 | 80 | 6 | 272.24 | 242.71 | 89.15 |
| 34 | S8-2 | 82 | 4 | 264.29 | 236.30 | 89.41 |
| 35 | S9-1 | 75 | 11 | 266.46 | 239.33 | 89.82 |
| 36 | S9-2 | 76 | 10 | 257.50 | 231.80 | 90.02 |
| 37 | S9-4 | 75 | 11 | 257.00 | 231.80 | 90.19 |

It is interesting to note that ideal $I_{sp,vac}$* for S1-2 occurs at no aluminum addition which is desirable due to the decrease in toxic fumes emitted because of aluminum. However, the crux of the paper is to compare the ideal and real starred specific impulses at vacuum.

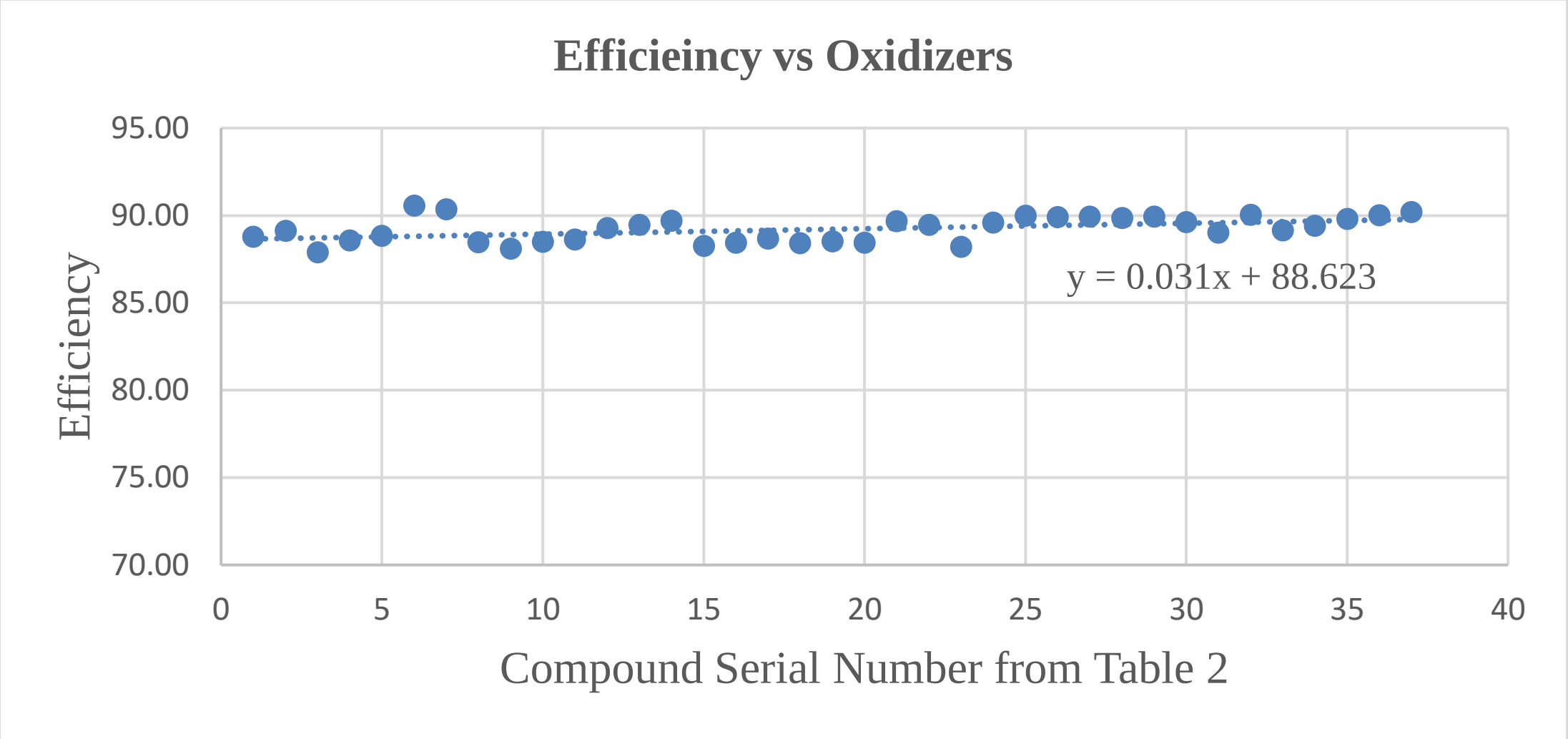

Figure 8. Various compounds and their efficiency

It is evident from Figure 8 that the real $I_{sp,vac}$* is a constant multiplied to the ideal $I_{sp,vac}$*. This paper, thus, provides a simple factor of 89.28% (median) as the multiplier to the ideal values from NASA CEA. The authors believe that the number of compounds presented in this paper are sufficient to provide strong proof for this trend, the factor can thus be used in vast compounds of said nature reducing computational efforts and time.

## 4. Conclusions

This paper presents a study of the propulsion performance of 37 compounds which show promising replacements to AP as an oxidizer. Supersonic nozzle flow simulations are performed using an opensource platform, OpenFOAM. Compared to ideal $I_{sp,vac}$* calculated using NASA CEA, boundary layer losses, turbulence losses and divergence losses get accounted for in the simulations. Grid convergence and independence check have been performed to ensure correct and fast results. The starred vacuum specific impulse compositions are compared with ideal values. The stated losses are seen to reduce the specific impulse to 88 - 91% of the ideal value. The computational tool presented here is handy for quick calculations to theoretically analyze the propulsion performance of newly synthesized compounds for propulsion applications.

## 5. Acknowledgements

The authors would like to acknowledge the grant from the Defense Research and Development Organization (Grant No: RD/0117-COPTS00-001) for supporting this work.